\documentclass{Interspeech2024}

\usepackage{algorithm}
\usepackage[noend]{algpseudocode}
\usepackage{amsmath}

\interspeechcameraready

\usepackage{multirow}
\title{ModalityMirror: Improving Audio Classification in Modality Heterogeneity Federated Learning with Multimodal Distillation}

\name[affiliation={1, *}]{Tiantian}{Feng}
\name[affiliation={1, *}]{Tuo}{Zhang}
\name[affiliation={1}]{Salman}{Avestimehr}
\name[affiliation={1}]{Shrikanth S.}{Narayanan}

\address{
    $^1$ Ming Hsieh Department of Electrical and Computer Engineering, University of Southern California
}
\email{tiantiaf@usc.edu, tuozhang@usc.edu}


\begin{document}

\maketitle
\def\thefootnote{*}\footnotetext{The first two authors contributed equally.}

\begin{abstract}
Multimodal Federated Learning frequently encounters challenges of client modality heterogeneity, leading to undesired performances for secondary modality in multimodal learning. It is particularly prevalent in audiovisual learning, with audio is often assumed to be the weaker modality in recognition tasks. To address this challenge, we introduce \texttt{ModalityMirror} to improve audio model performance by leveraging knowledge distillation from an audiovisual federated learning model. \texttt{ModalityMirror} involves two phases: a modality-wise FL stage to aggregate uni-modal encoders; and a federated knowledge distillation stage on multi-modality clients to train an unimodal student model. Our results demonstrate that \texttt{ModalityMirror} significantly improves the audio classification compared to the state-of-the-art FL methods such as \texttt{Harmony}, particularly in audiovisual FL facing video missing. Our approach unlocks the potential for exploiting the diverse modality spectrum inherent in multi-modal FL.
\end{abstract}


\section{Introduction}
Multi-modal sensing systems are increasingly important in a range of real-world applications, such as human activity recognition, health monitoring, and emotion recognition~\cite{Feng2023FedMultimodalAB, Liu2023EffectivenessOV, Awada2023ANP, liu2023development, Zhang2024CreatingAL, Zhang2024CreatingAL}. While these multimodal systems provide promises in enriching human experiences, they have to address one fundamental challenge: achieving robust recognition without compromising data privacy, especially when data is collected in sensitive environments such as hospitals, schools, and homes. Specifically, Federated Learning (FL) has been recently introduced as a privacy-preserving machine learning paradigm. It enables model training on a central server using shared model parameters from edge devices, eliminating the need to transfer local data~\cite{Zhang2022FederatedLF}.

Despite its potential to protect privacy, FL frequently faces critical challenges in achieving competitive performances when encountering device and data heterogeneity. Devices in FL systems vary widely in their capabilities and the modalities they can collect. A significant aspect of this heterogeneity is the variance in data modalities: some devices learn rich information by processing multi-modal data, while others are constrained to use single-modal data. This variation introduces a much less explored challenge in FL, known as modality heterogeneity. Specifically, this challenge may frequently occur in applications involving audio-visual recognition. Notably, visual data, such as videos or images, carry sensitive information about a person, including facial geometric, body shapes, and other biometric data. This prevents many service providers from collecting, storing, and processing visual information, even in FL.

Furthermore, particularly within audiovisual recognition, the dynamics between dominant and weak modalities play a critical role in recognition performance. Typically, visual input is considered the dominant modality in audiovisual recognition, heavily influencing model training due to its rich, detailed content. In contrast, the audio modality is often recognized as a weaker role in audiovisual recognition. This phenomenon could be be more significant in FL with heterogeneous modalities described earlier, with some devices of multi-modal data and others of audio data, leading to decreased audio classification. 
Existing literature in addressing modality heterogeneity ~\cite{Ouyang2023HarmonyHM, Chen2022FedMSplitCF} has primarily focused on increasing the utilization of single-modality local data for multi-modality training. However, such methods typically do not address the limitations of single-modality clients, who cannot practically employ multi-modal models.
Therefore, our research specifically aims to improve the classification performance of clients limited to audio data, the conventionally weaker modality in audio-visual recognition, within FL with heterogeneous modalities.

In response, we introduce \texttt{ModalityMirror}, a novel FL paradigm to tackle modality heterogeneity in FL involving audiovisual recognition. The primary object of \texttt{ModalityMirror} is to enhance audio classification through knowledge distillation from the audiovisual model trained by FL. The \texttt{ModalityMirror} involves two distinct training stages: a modality-wise FL stage to learn the audiovisual model; a federated knowledge distillation occurs among multi-modality clients to train an audio student model. Empirical results demonstrate that \texttt{ModalityMirror} yields significant improvements in accuracy, especially in the audio-modality models when compared to recent state-of-the-art FL frameworks such as \texttt{Harmony}~\cite{Ouyang2023HarmonyHM}. 
Furthermore, ablation studies reveal that \texttt{ModalityMirror} is particularly effective at enhancing the audio model's performance on labels where the audio modality alone provides insufficient information for classification.

\begin{figure*}[t]
	\centering
	\includegraphics[width=\linewidth]{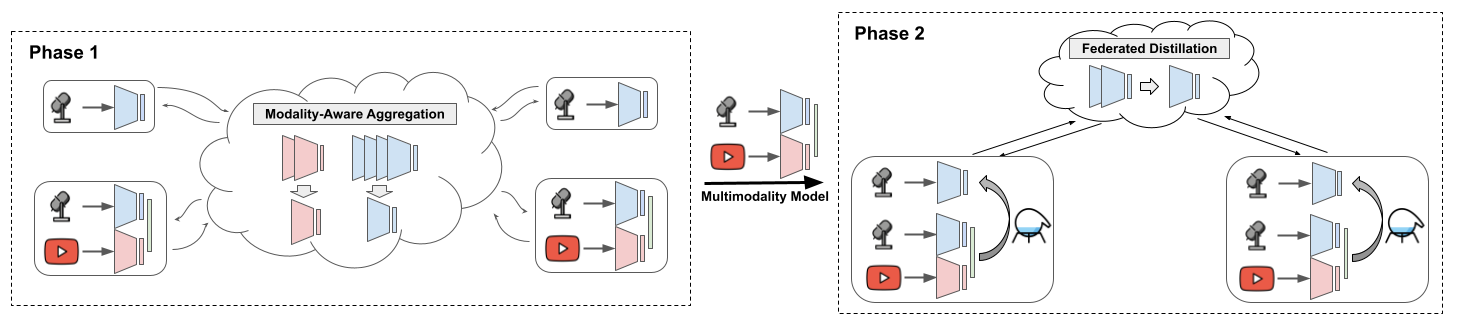}
    \vspace{-6mm}
    \caption{System Architecture of \texttt{ModalityMirror}. \texttt{ModalityMirror} comprises two distinct stages: modality-aware federated learning across all nodes, followed by federated distillation leveraging data-rich nodes.}
    \label{fig:gpt-mm}
    \vspace{-3mm}
\end{figure*}

\section{Related Study}
\textbf{Status Quo and Their Limitations.}
In the domain of multi-modal federated learning (FL), frameworks such as \texttt{CreamFL}~\cite{Yu2023MultimodalFL} and \texttt{Harmony}~\cite{Ouyang2023HarmonyHM} have emerged as significant contributions, albeit with notable limitations in real-world scenarios involving missing modalities. \texttt{CreamFL} utilizes a contrastive representation-level ensemble to construct a comprehensive server model from heterogeneous multi-modal client data, but its reliance on a quality-dependent public dataset for server-side training and aggregation poses practical challenges in diverse settings where such datasets may be unavailable. 
\texttt{Harmony} introduces a two-stage framework involving modality-specific federated learning and federated fusion, effectively segmenting multi-modal network training. However, it inadequately addresses missing modalities and predominantly outputs multi-modality models, thereby excluding single-modality clients and limiting its practical applicability.

\vspace{-2mm}
\begin{algorithm}
    \caption{Modality-aware FL Training}
    \begin{algorithmic}[1]
        \State{\textbf{Initialize: }}{$\mathbf{\theta^{A}}$, $\mathbf{\theta^{V}}$}
        
        \State{\textbf{Server executes:}}
        \For{Each round $t=0,...,T-1$}
            \State{Sample Audio Modality Clients $\mathcal{S}^{A}$}
            \State{Sample Multimodal Clients $\mathcal{S}^{M}$}
            
            \For{Each client $p \in \mathcal{S}^{A}$ in parallel}
            
                \State $\mathbf{\theta_{p}^{A}} \gets \mathbf{\theta^{A}}$
                \State $\mathbf{\theta_{p}^{A}} \gets$ ClientLocalTraining($\mathbf{\theta_{p}^{A}}$, $\mathcal{D}_{p}^{A}$)
                
            \EndFor

            \For{Each client $q \in \mathcal{S}^{M}$ in parallel}
            
                \State $\mathbf{\theta_{q}^{A}}, \mathbf{\theta_{q}^{V}} \gets \mathbf{\theta^{A}}, \mathbf{\theta^{V}}$
                \State $\mathbf{\theta_{q}^{A}}, \mathbf{\theta_{q}^{V}} \gets$ ClientLocalTraining($\mathbf{\theta_{q}^{A}}$, $\mathbf{\theta_{q}^{V}}$, $\mathcal{D}_{q}^{M}$)
                
            \EndFor
            
            \State $\mathbf{\theta^{V}} \gets \frac{1}{|\mathcal{S}^{M}|} \sum_{q\in \mathcal{S}^{M}} \mathbf{\theta_{q}^{V}}$

            \State $\mathbf{\theta^{A}} \gets \frac{1}{|\mathcal{S}^{A}|+|\mathcal{S}^{M}|} (\sum_{p\in \mathcal{S}^{A}} \mathbf{\theta_{p}^{A}}+\sum_{q\in \mathcal{S}^{M}} \mathbf{\theta_{q}^{A}}$)
            
        \EndFor
        
    \end{algorithmic}
    \label{alg:modality_aware_fl}

\end{algorithm}
\vspace{-4mm}

\section{Method}
\subsection{Problem Formulation}
Here, we describe the problem formulation of our multi-modal FL with modality heterogeneity. Specifically, we refer to multimodal clients as data-rich clients.  First, we denote client size as $N$ in FL.  Given our focus on audio-visual recognition, we define the audio and visual modalities as $A$ and $V$, respectively. Moreover, as we emphasize studying the visual modality missing in modality heterogeneity FL, we introduce $r$ as the ratio of clients with missing visual modality, leading to the total number of audio-only modality clients as $N^{A}=rN$. Simultaneously, the number of multi-modal clients is denoted as $N^{M}=(1-r)N$. For each audio modality client $p\in[N^{A}]$, the associated audio modality dataset is defined as $\mathcal{D}_{p}^{A} = \{x_{{i}}^{A}, y_{i}\}$, where $i\in\mathbb{N}$. Additionally, each audio modality client $p\in[N^{A}]$ is assumed to adopt the audio model $\theta_{p}^{A}$ trained on $\mathcal{D}_{p}^{A}$. Similarly, for each multi-modal client $q\in[N^{M}]$, the multi-modal data is represented as $\mathcal{D}_{q}^{M} = \{x_{i}^{A}, x_{i}^{V}, y_{i}\}$, and the associated model is denoted as $\theta_{q}^{M}$=$\{\theta_{q}^{A}$, $\theta_{q}^{V}\}$, where $q\in[N^{M}]$.

\begin{algorithm}
    \caption{Federated Distillation}
    \begin{algorithmic}[1]
        \State{\textbf{Initialize From Modality-aware Training: }}{$\mathbf{\theta^{M}}$, $\mathbf{\theta^{A}}$}
        
        \State{\textbf{Server executes:}}
        \For{Each round $t=0,...,T-1$}
            \State{Sample Multimodal Clients $\mathcal{S}^{M}$}
            
            \For{Each client $q \in \mathcal{S}^{M}$ in parallel}
            
                \State $\mathbf{\theta_{q}^{A}} \gets \mathbf{\theta^{A}}$

                \For{Batch Data $x^{A}, x^{V}, y$ from $\mathcal{D}_{q}^{M}$}

                    \State $p_{i}^{A} \gets \mathcal{F}(\theta_{q}^{A}, x^{A})$

                    \State $p_{i}^{M} \gets \mathcal{F}(\theta^{M}, x^{A}, x^{V})$

                    \State $\mathbf{\theta}_{q}^{A} \gets \mathbf{\theta}_{q}^{A}-\eta\nabla_{\theta}(\mathcal{L}_{ce}(p^{A}, y) + $ $KL(p^{A}||p^{M})$
                
                \EndFor
                
            \EndFor
            
            \State $\mathbf{\theta^{A}} \gets \frac{1}{|\mathcal{S}^{A}|} \sum_{q\in \mathcal{S}^{M}} \mathbf{\theta_{q}^{A}}$

        \EndFor
        
    \end{algorithmic}
    \label{alg:fl_distillation}
    
\end{algorithm}

\subsection{\texttt{ModalityMirror} For Audio-visual Recognition}
\subsubsection{Overview}
Our proposed \texttt{ModalityMirror} is motivated by two principal objectives: first, to mitigate the challenge of modality heterogeneity within FL; second, to enhance the performance of devices that support only a weaker data modality, such as the audio modality in this work, by leveraging the knowledge from multimodal clients within the FL training. Consequently, we present a two-phase framework comprising modality-aware Federated Learning followed by Federated Distillation. The distillation training aims to decouple multi-modal and single-modal training processes. Figure~\ref{fig:gpt-mm} illustrates the \texttt{ModalityMirror} system architecture. In the following, we describe the details of each training stage.

\vspace{-1mm}
\subsubsection{Modality-Aware Federated Learning.} \label{fl}
\vspace{-1mm}
We present the modality-aware Federated Learning in Algorithm~\ref{alg:modality_aware_fl}, which enables collaborative training across nodes with distinct modalities.
In this phase, nodes that support multi-modal data trains a multi-modal model locally. In contrast, nodes constrained to a single modality, like the audio modality, focus on training a model specific to their available modality. In this phase, the server applies a modality-aware aggregation to mitigate the disparities introduced by modality heterogeneity. 
%

This modality-aware FL is practical, given that most multimodal models, including audio-visual recognition models, adopt the dual-encoder architecture with each encoder tied to a specific modality. The embeddings from the modality-specific encoder are concatenated and fed into the classification layers for audio-visual recognition. This embedding-level fusion allows the server to execute a nuanced aggregation process, accommodating nodes with distinct modalities by selectively aggregating encoder weights to refine the global model.

\begin{table*}[t]
    \centering
    \caption{Accuracy performance across benchmark datasets for audio-only modality under varying video modality missing rate.}
    \label{tab:main}
    \resizebox{1.0\textwidth}{!}{
    \begin{tabular}{ccccccc}
        \toprule
        \multirow{2}{*}{\textbf{Dataset}} & \multirow{2}{*}{\textbf{Method}} & \multicolumn{5}{c}{\textbf{Video Modality Missing Rate}} \\
        \cmidrule(lr){3-7}
        & & \textbf{10\%} & \textbf{20\%} & \textbf{30\%} & \textbf{40\%} & \textbf{50\%} \\
        \cmidrule(lr){1-1} \cmidrule(lr){2-2} \cmidrule(lr){3-7}
        \multirow{4}{*}{\textbf{UCF101}} & \textbf{UniFL} & 31.28 (\textpm\;0.08)& 32.92 (\textpm\;1.78) & 28.96 (\textpm\;0.97) & 29.63 (\textpm\;0.26)  & 28.03 (\textpm\;2.15)  \\
        & \textbf{MultiFL} & 28.08 (\textpm\;1.77) & 31.11 (\textpm\;0.63) & 31.79 (\textpm\;1.26) & 33.49 (\textpm\;1.62) & 32.52 (\textpm\;2.83) \\
        & \texttt{Harmony} & 36.93 (\textpm\;0.60) & 36.93 (\textpm\;0.60) & 36.93 (\textpm\;0.60) & 36.93 (\textpm\;0.60) & 36.93 (\textpm\;0.60) \\
        & \textbf{\texttt{ModalityMirror}} & \textbf{41.77} (\textpm\;0.30) & \textbf{40.12} (\textpm\;0.43) & \textbf{40.43} (\textpm\;0.64) & \textbf{39.51} (\textpm\;0.81) & \textbf{40.48} (\textpm\;0.37) \\
        \cmidrule(lr){1-1} \cmidrule(lr){2-2} \cmidrule(lr){3-7}
        \multirow{4}{*}{\textbf{ActivityNet}} & \textbf{UniFL} & 6.97 (\textpm\;0.04) & 6.98 (\textpm\;0.05) & 7.45 (\textpm\;0.06) & 6.14 (\textpm\;0.04) & 6.79 (\textpm\;0.03) \\
        & \textbf{MultiFL} & 11.19 (\textpm\;0.10) & 11.70 (\textpm\;0.11) & 11.91 (\textpm\;0.05) & 12.55 (\textpm\;0.28) & 13.09 (\textpm\;0.13) \\
        & \texttt{Harmony} & 14.35 (\textpm\;0.49) & 14.35 (\textpm\;0.49) & 14.35 (\textpm\;0.49) & 14.35 (\textpm\;0.49) & 14.35 (\textpm\;0.49) \\
        & \textbf{\texttt{ModalityMirror}} & \textbf{15.82} (\textpm\;0.02) & \textbf{15.10} (\textpm\;0.05) & \textbf{15.75} (\textpm\;0.01) & \textbf{15.05} (\textpm\;0.03) & \textbf{15.63} (\textpm\;0.01) \\
        \bottomrule
    \end{tabular}
    }
\end{table*}

\subsubsection{Federated Distillation}

%

In the second phase, our objective is to enhance the performance of audio-only clients by leveraging knowledge from data-rich (multi-modal) clients.
Specifically, we introduce Federated Distillation training to unify model knowledge acquired from clients with diverse modalities. This process particularly benefits audio-modality nodes by enabling them to distill and learn multimodal knowledge from the audio-visual model initially trained by all nodes. The complete Federated Distillation algorithm is presented in Algorithm~\ref{alg:fl_distillation}.

In practice, data-rich clients perform the distillation process locally, training an audio-modality model that distills knowledge from the audio-visual model. The distillation loss used in the local training is the KL divergence $KL(p^{A}||p^{M})$ between the multimodal output $p^{M}$ and the audio model output $p^{A}$. Here, we introduce the temperature value $T$ to adjust the entropy of the predicted softmax probabilities by $p^{A} \propto exp(\frac{\log(p^{A})}{T})$ and $p^{M} \propto exp(\frac{\log(p^{M})}{T})$. This distilled audio-modality model is then applied with federated aggregation, resulting in an audio-modality model that incorporates the collective knowledge of both audio and audio-visual data sources. This federated distillation phase thus ensures that all participating nodes, regardless of their data modality, benefit from the enriched learning landscape offered by the multi-modality model.


\section{Datasets and Experimental Details}
\subsection{Audio-visual Datasets}
In this work, we select two popular multimedia action recognition datasets, UCF101~\cite{Soomro2012UCF101AD} and ActivityNet~\cite{Heilbron2015ActivityNetAL}, to assess the \texttt{ModalityMirror} framework for audio-visual recognition. We refer the reader to ~\cite{Feng2024CanTM} for detailed label distribution within these datasets. Motivated by studies presented in ~\cite{Feng2024CanTM, Gong2022ContrastiveAM}, we extract the video frame in the middle of the video as the visual modality. We conduct 3-fold cross-validation following the standard fold splits in training the UCF101 dataset. Increasingly, we apply the standard train/validation/test split in experimenting with the ActivityNet dataset. In line with ~\cite{Feng2023FedMultimodalAB}, we partition both datasets into 100 distinct data silos using a Dirichlet distribution with $\alpha=0.1$, simulating an extreme non-IID setting. More detailed about datasets are listed in Appendix~\ref{sec:dataset}.



\subsection{Model Details}
We utilize pre-trained ResNet-18~\cite{He2015DeepRL} and SSAST~\cite{Gong2021SSASTSA} models as the backbones for modeling visual and audio modalities, respectively. The SSAST model is initialized with pre-trained weights as distributed in \cite{Gong2021SSASTSA}. For ResNet-18, we employ the ImageNet-based pre-trained weights available through PyTorch's torch-vision library (version 0.4.1). As delineated in Section~\ref{fl}, late fusion is used as the principal method for fusing different modalities.

\subsection{Evaluation Baselines}

\noindent \textbf{\texttt{Harmony}} ~\cite{Ouyang2023HarmonyHM} is the first selected baseline, achieving state-of-the-art performance in heterogeneous multi-modal FL. \texttt{Harmony} adopts a two-stage framework. Initially, all participating nodes collectively train two distinct unimodal encoders within the FL paradigm. In the second stage, only multi-modal nodes engage in the collaborative training of the multimodal model, with the initialization of feature encoders using the model weights refined during the first stage, followed by fine-tuning. The unimodal encoders trained in the first stage would be used for the single-modality clients. 

\vspace{0.3mm}
\noindent \textbf{\texttt{UniFL}}~\cite{Xiong2022AUF}: 
Motivated by \texttt{Harmony}, we include one naive heterogeneous multi-modal FL baseline, \texttt{UniFL}~\cite{Xiong2022AUF}, in comparison. \texttt{UniFL} represents a scenario in which nodes sharing identical data modalities collaboratively learn a model. For example, nodes with audio modality would collaboratively train an audio model, while nodes with multi-modalities are learning an audio-visual model collectively.

\vspace{0.3mm}
\noindent \textbf{\texttt{MultiFL}}~\cite{Zhao2021MultimodalFL}: Apart from \texttt{UniFL}, we experiment with \texttt{MultiFL} described in \cite{Zhao2021MultimodalFL}. In \texttt{MultiFL}, all nodes participate in the training, and the server returns the unimodal encoders by averaging the uni-modal and multi-modal nodes.

\begin{figure*}[h!]
        \vspace{-3mm}
	\centering
	\includegraphics[width=0.9\linewidth]{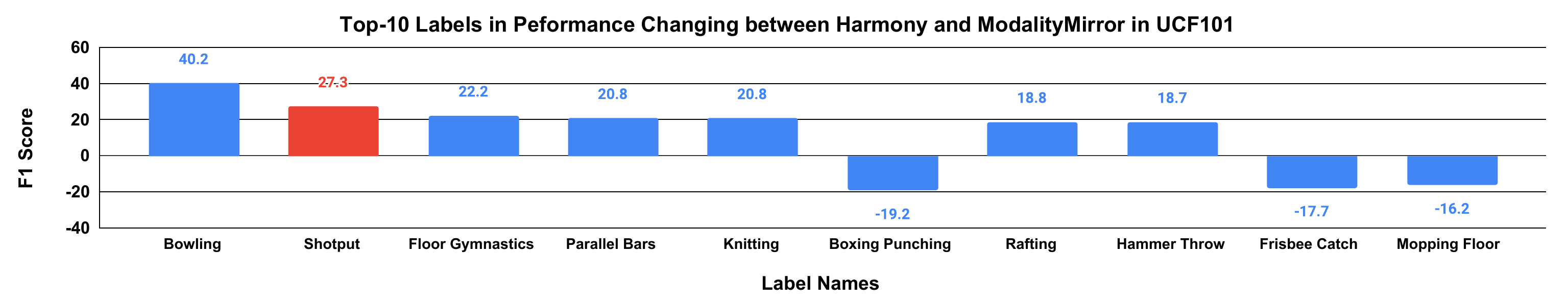}
        \includegraphics[width=0.9\linewidth]{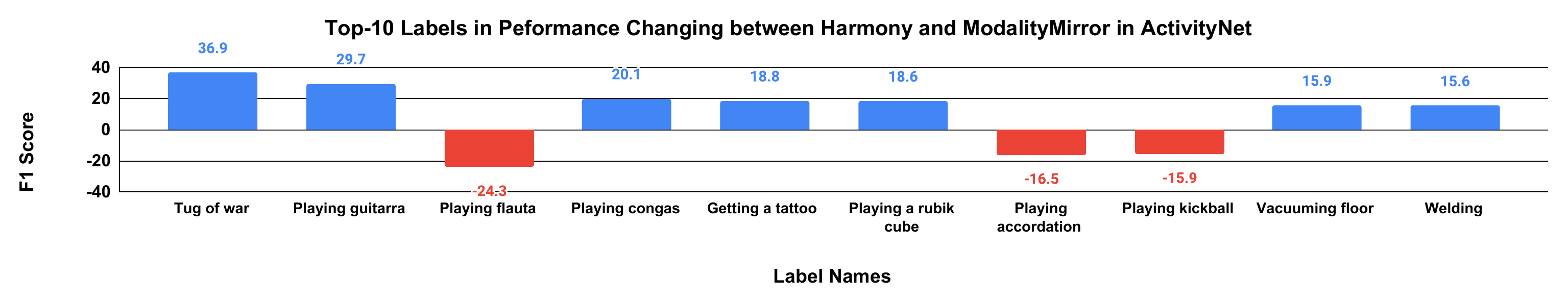}
    \vspace{-3mm}
    \caption{Comparative Analysis of Top 10 relative performance changes across individual labels between \texttt{Harmony} and \texttt{ModalityMirror} with UCF101 and ActivityNet dataset in audio modality respectively, where each bar represents the difference in F1 scores for a specific label.
    }
    \label{fig:ucf101}
    \vspace{-3mm}
\end{figure*}

\subsection{Evaluation Metrics}
As this paper concentrates on the performance of clients who only hold weaker (audio) modalities for training, we use the top-1 test accuracy of the audio-modality model of UCF101 as our evaluation metric. Due to relatively low performance on ActivityNet, we report the top-5 test accuracy. We ran the experiments with three distinct seeds and reported the average performance and variance on both datasets.

\section{Experiments}
\subsection{End-to-End Performance}
We begin by comparing the end-to-end performance of \texttt{ModalityMirror} on benchmark datasets. For consistency, all experiments conducted in Table~\ref{tab:main} follow a uniform setting of 200 communication rounds and 1 local epoch. In each round, the server randomly samples 10 clients for local training across both datasets. We searched the client learning rate in the range of 1.00E-05 to 1.00E-01, and selected 5.00E-04 for experiments. We implement the FedAvg as the aggregation function.

Table~\ref{tab:main} summarizes our results, where we make three key observations: (1) Across varying video modality missing rates from 10\% to 50\%, \texttt{ModalityMirror} consistently outperforms other baselines in accuracy on both datasets.
(2) \texttt{MultiFL} marks the performance of Modality-aware Federated Learning as detailed in Algorithm~\ref{alg:modality_aware_fl}. The increment from \texttt{ModalityMirror} to \texttt{MultiFL} underscores the effectiveness of Federated Distillation, as described in Algorithm~\ref{alg:fl_distillation}.
(3) \texttt{Harmony} indicates the performance that only relies on the audio data for training. The comparative advantage of \texttt{ModalityMirror} over \texttt{Harmony} suggests that though it only takes a single modality as the input, the audio model could benefit from distilling knowledge from multi-modality model.


\subsection{Impact of Video Modality Missing Rate}
\begin{figure}[h!]
    \vspace{-3mm}
	\centering
	\includegraphics[width=\linewidth]{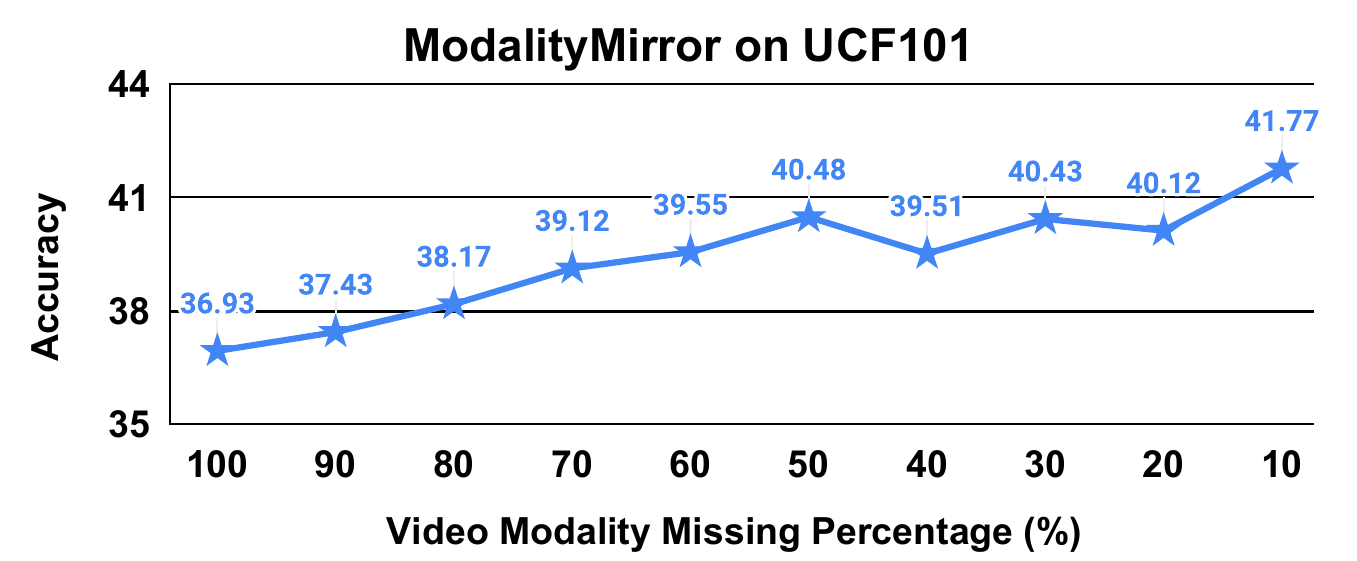}
    \vspace{-3mm}
    \caption{Accuracy performance of \texttt{ModalityMirror} on UCF101 under various video modality missing rates.}
    \label{fig:rate}
    \vspace{-3mm}
\end{figure}

In order to investigate the impact of video modality missing rate on \texttt{ModalityMirror}, we test \texttt{ModalityMirror} with video modality missing rate ranging from 10\% to 90\% on the UCF101 dataset. As the results shown in Figure~\ref{fig:rate}, the performance of \texttt{ModalityMirror} increases as the video missing rate decreases, indicating that a positive correlation between the completeness of video data and the model's performance.

\subsection{Understanding the Advantages of \texttt{ModalityMirror}}
To elucidate the benefits of \texttt{ModalityMirror}, we analyze the F1 score performance differential for individual labels between \texttt{ModalityMirror} and \texttt{Harmony} across both datasets under a 30\% video modality missing scenario. Figure~\ref{fig:ucf101} highlights the top 10 labels exhibiting significant relative performance changes. We make two key observations:
(1) Through federated distillation from multi-modality models, \texttt{ModalityMirror} notably enhances the audio model's performance on labels where the audio modality alone provides limited information. For instance, in the UCF101 dataset, \texttt{Harmony} achieves an F1 score of 9.12 for the \textit{bowling} class and 0.0 for \textit{Tug of War} in ActivityNet, reflecting the insufficiency of audio cues in these videos for accurate classification. The integration of knowledge from the image encoder via model distillation allows the audio model to better classify these challenging categories. While this approach occasionally reduces performance on labels predominantly reliant on audio, such as \textit{playing flute} and \textit{playing accordion}, the overall enhancement in performance, as evidenced in Table~\ref{tab:main}, demonstrates a net benefit.
(2) \texttt{ModalityMirror} significantly reduces classification ambiguity in audio-based tasks. For certain labels, the distinction between audio features is minimal; for instance, \textit{Playing congas} and \textit{Playing drums} in ActivityNet exhibit highly similar audio profiles. Consequently, \texttt{Harmony} attains F1 scores of 33.02 and 19.42 for these categories, respectively. In contrast, by leveraging insights from the multi-modality model, \texttt{ModalityMirror} is able to more effectively differentiate between these two labels, enhancing their performance to F1 scores of 53.10 and 26.24, respectively.

\section{Conclusion and Limitation}
We propose \texttt{ModalityMirror}, a distillation-based FL framework aiming to improve audio classification in modality heterogeneity FL.
Our experimental results showcase the competitive performance of \texttt{ModalityMirror} when compared to state-of-the-art FL methods. Moreover, through detailed ablation studies, we demonstrate that \texttt{ModalityMirror} notably enhances the audio model’s performance on labels where the audio modality alone provides limited information. 

\noindent \textbf{Limitation and future work.} The empirical results indicate that \texttt{ModalityMirror} may occasionally reduce performance on labels associated with strong acoustic characteristics. We suspect that the performance reduction is mainly due to the federated distillation process unlearns specific representative acoustic features during the distillation of multimodal knowledge. Therefore, in future works, we plan to investigate the root cause of accuracy drop for audio-specific labels and explore alternative learning objective to minimize information loss.

\bibliographystyle{IEEEtran}
\bibliography{mybib}

\newpage
\appendix
\section{Problem Formulation}
Without loss of generality, we denote our multi-modal FL with modality heterogeneity with a client size of $N$. Specifically, we also refer to multimodal clients as data-rich clients. Given our specific focus on audio-visual applications, we define the audio and visual modalities as $A$ and $V$, respectively. Moreover, as we emphasize studying the visual modality missing in modality heterogeneity FL, we introduce $r$ as the ratio of clients with missing visual modality, leading to the total number of audio-only modality clients as $N^{A}=rN$. Simultaneously, the number of multi-modal clients is denoted as $N^{M}=(1-r)N$. For each audio modality client $p\in[N^{A}]$, the associated audio modality dataset is defined as $\mathcal{D}_{p}^{A} = \{x_{{i}}^{A}, y_{i}\}$, where $i\in\mathbb{N}$. Additionally, each audio modality client $p\in[N^{A}]$ is assumed to adopt the audio model $\theta_{p}^{A}$ trained on $\mathcal{D}_{p}^{A}$. Similarly, for each multi-modal client $q\in[N^{M}]$, the multi-modal data is represented as $\mathcal{D}_{q}^{M} = \{x_{i}^{A}, x_{i}^{V}, y_{i}\}$, and the associated model is denoted as $\theta_{q}^{M}$=$\{\theta_{q}^{A}$, $\theta_{q}^{V}\}$, where $q\in[N^{M}]$. In modality heterogeneity FL, we aim to obtain the global multi-modal model $\theta^{M}$. 
The clients cooperate in training parameters $\theta_{q}^{M}$ and $\theta_{p}^{A}$ with the aim of solving the optimization problem formulated as follows:
\begin{equation}
    \min\limits_{\theta_{p}^{A}} f(\theta_{p}^{A}) := \sum_{p\in [N^{A}]} w_p \mathcal{F}_p(\theta_{p}^{A}) 
\end{equation}
\begin{equation}
    \min\limits_{\theta_{q}^{M}} f(\theta_{q}^{M}) := \sum_{q\in [N^{M}]} w_q \mathcal{F}_q(\theta_{q}^{A}, \theta_{q}^{V}) 
\end{equation}
where $\mathcal{F}_p(\cdot)$ and $\mathcal{F}_q(\cdot)$ represents the local objective of client $p$ and $q$, respectively. $w_p$ and $w_q$ denotes the aggregation weight of client $p$ and $q$, satisfying $w_p \geq 0$, $w_q \geq 0$ and $\sum_{p\in [N^{A}]} w_p=1$, $\sum_{q\in [N^{M}]} w_q=1$.

\section{Audio-visual Datasets} \label{sec:dataset}

\noindent \textbf{UCF101} dataset focuses on human action recognition, including videos from 101 sport-related action categories. The majority of the video is sourced from YouTube, leading to a total of more than 10k videos. However, we identified that videos with only 51 categories include both video and audio modalities. This leads to fewer than 7,000 videos for the experiments. As the video duration in this dataset varies from a few seconds to over 30 seconds, we decided to constrain the input audio duration to 5 seconds. In line with ~\cite{Feng2023FedMultimodalAB}, we partition this dataset into 100 distinct data silos using a Dirichlet distribution with $\alpha=0.1$, aiming to replicate an extreme non-IID setting.

\noindent \textbf{ActivityNet} is a similar dataset to UCF101 to study human action recognition but includes human actions in a broader spectrum. The complete dataset has 203 unique human action classes for everyday life, while video data with only 200 human actions are presented with both audio and visual modalities. This reduces the total data to 18,976 instances. We adopt the same partition strategy as with UCF101, dividing ActivityNet into 100 independent data silos via a Dirichlet distribution with $\alpha=0.1$ to mirror an extreme non-IID scenario.

\end{document}